# Inter-cellular Interactions and Patterns: Vertebrate Development and Embryonic Stem Cells

Eric D. Siggia

*Rockefeller University: Address*
*1230 York Avenue New York, NY 10065 USA*
*E-mail siggiae@rockefeller.edu*

Development from egg to embryo to adult is a fascinating instance of biological self-organization for which genetics has supplied us with a parts list. It remains to find the principles organizing the assembly of those parts. In the last decade embryonic stem cells (ESC) have provided the material from which to build the mammalian embryo. This review, for a quantitative audience, explains why colonies of ESC are an ideal system with which to peal back the multiple layers of regulation that make embryonic development such a robust process. It formed the basis of a presentation at the 27$^{th}$ Solvay Conference on the Physics of Living Matter edited by Boris Shraiman.

## 1. Introduction:

It is foolish to summarize a subject as vast as vertebrate development, yet a more focused discussion would sacrifice the bits of generality I will try to convey. If physicists are fond of 'self organization' and 'symmetry breaking' then biology offers no more dramatic example than embryology. It puts to shame any of the contrived systems invented for systems biology; real physiology remains more interesting. The reader looking for the universal theory uniting just some of the topics in our session: biofilms, flocking behavior, and development should look elsewhere. Attempting to treat them together leads to a degree of superficiality that illuminates nothing. Slogans that biology, is robust, modular, evolvable, etc., are too vague to be useful.

These remarks are aimed towards the student of biology from the mathematical and physical sciences, who wishes for a few provisional guideposts as to what problems seem most approachable at the current instant. In almost all cases, autonomous first principles theory is a fool's errand. It would appear to outsiders that biological data is infinite (e.g., there are upwards of 20,000 papers in Pubmed that mention each of the six or so intercellular signaling pathways that pattern the early vertebrate embryo), yet it has been the experience of most in the field, that theoretical ideas require new data. So this review aims to provide the skeleton of concepts that could motivate the next round of experiments, and highlight the systems most likely to provide answers.

In searching for principles, why study vertebrates and not arthropods; all the signaling pathways are present in arthropods, without the huge degeneracy of components. Genetics is easier, and evolution moves more quickly and has created fascinating variety, (see remarks of Nipam Patel). But there is a natural interest in our selves, common interests mean more shared reagents, techniques cell lines, and it's not a sin to be medically relevant. But the real advance that makes vertebrate development interesting for the quantitative class is pluripotent stem cells specifically in what follows human embryonic stem cells, hESC. These cells quite literally give rise to all cells of the adult. Basic cell culture taught us about intracellular signaling and organelles (see the report of Lippincott-Schwartz) exploiting what are basically cancer cells, HeLa [1]being the most notorious example. Such systems are a very dubious starting point for problems of cell communication and embryology, even if one can engineer them with some of the right constituents. Biological components do many things in-vitro that do not happen in-vivo. The same caveat applies to stem cells and at crucial points an embryological comparison is needed, but in the appropriate context stem cells do the appropriate thing, as shown by the canonical grafting experiments.

## 2. Gastrulation

A favorite system for experimental vertebrate embryology from the early 20[th] century is the frog *Xenopus*. The eggs are 1.2mm in diameter, they are easily fertilized on demand in the lab, and become swimming tadpoles in 2 days. No special regents needed, just pond water. The reader is invited to view one of the gastrulation movies on Xenbase or YouTube. The egg begins with top and bottom (animal, vegetal) hemispheres distinguished, sperm entry defines the future dorsal side. Signals from the vegetal side, induce a band of mesoderm cells around the equator from the multipotent animal cap (hemisphere) cells. At gastrulation this band closes like a purse string, by converging towards the dorsal side. The converging cells dive under the epidermis, and form a stiff bundle, the notochord, that elongates and literally builds the anterior-posterior axis. The vegetal hemisphere is pulled inside and the cavity formed from the outside inward by the so-called convergence-extension movements becomes the future gut (the online movies essential here). The master of *Xenopus* gastrulation is Ray Keller at University of Virginia and his papers provide the best description we have for the forces driving these morphogenic movements e.g., [2]

While the embryo is dramatically changing shape, it also is laying down, very literally, a coordinate system defined by the HOX genes along the anterior-posterior axis. That morphogenesis and fate assignment happen simultaneously is quite essential, since the cues for position come precisely from the cell movements. The HOX genes are located in contiguous cluster in the genome and are expressed sequentially in time in the converging mesoderm band by very complex regulation tied to their genomic organization (see papers of D. Duboule Lausanne). The HOX expression is locked down when a cell goes through the point of convergence on the future dorsal side, the Spemann organizer, (see Wikipedia). Thus a temporal signal is converted into a spatial coordinate as the embryo builds its' body axes, Figure 1 [3], [4]. The organizer should not be thought of as defined structure like the gut, but rather a reaction center through which cells transit and change state. Although the organizer can be surgically transplanted to induce a second body axis, in the chick it can also regenerate following excision [5]. Exactly how the juxtaposition of tissues surrounding the organizer recreates the organizer is not understood, though recently an ectopic organizer was created in the chick by transplanting a patch of cells derived from hESC (preprint Martyn, Kannno, Ruzo, Siggia, Brivanlou)

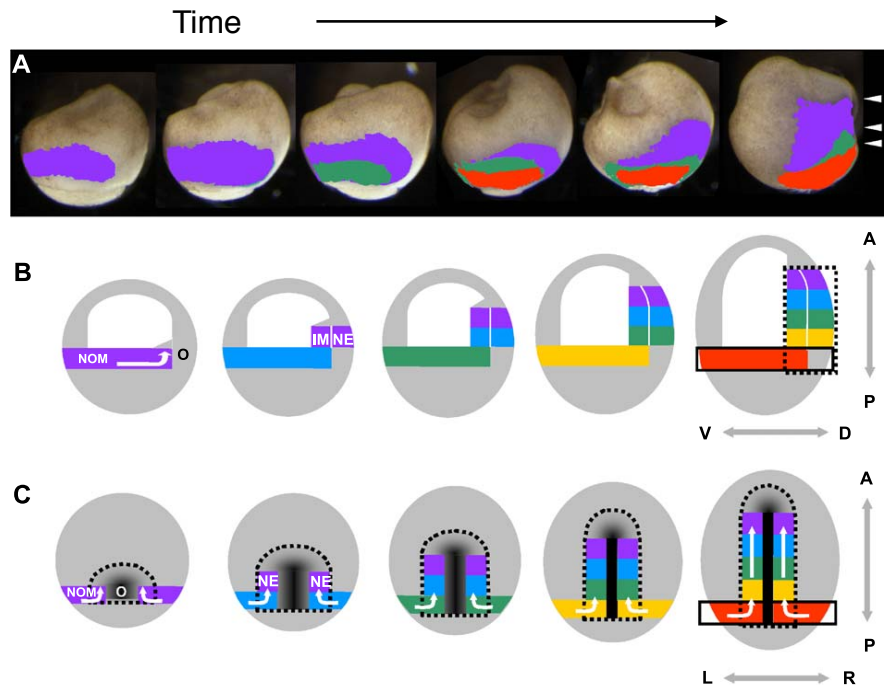

Figure 1 Temporal progression of HOX gene expression in the equatorial mesoderm is locked down on the anterior-posterior (A,P) axis. Sagittal sections are shown on the top two rows and a dorsal view on the bottom (V,D ventral, dorsal; L,R left,right). From [4], Figure 6.

The dorsal-ventral axis is established by the signaling pathways that recur through out development BMP, Nodal, WNT, and FGF/MAPK. They are very dynamic prior to gastrulation , Figure 2 [6] and more so afterwards, and it would be perilous to approximate the embryo as one dimensional in such circumstances [7]. The data in Figure 2 is derived by sectioning *Xenopus* embryos and staining the slices with antibodies for the transcriptions factors that move to the nucleus in response to the signals. Thus one records the net effect of the secreted morphogens and their inhibitors in the embryo. One might have hoped for more modern data from light sheet microscopy on the transparent zebra fish embryo, but as of this writing nothing comparable in scope to the 2002 Schohl paper is available. Modern technology consumes its creators.

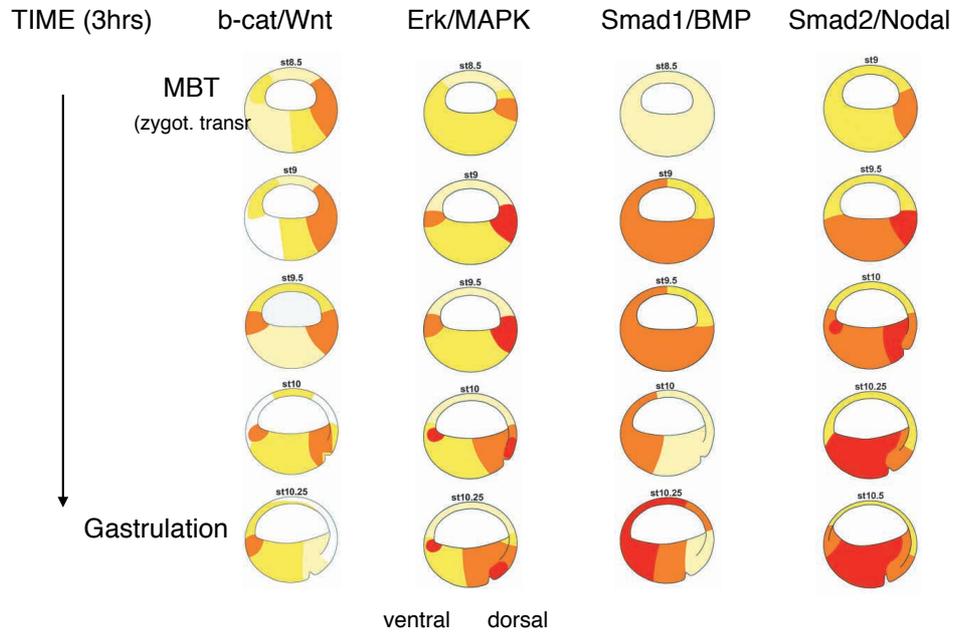

Figure 2. Sagittal views of the activity of the canonical signaling pathways just prior to gastrulation in Xenopus. from [6], Fig. 9,10. MBT or mid blastula transition denotes the beginning of general zygotic transition followed 3hr later by gastrulation. The color scale for intensity places red highest and yellow lowest.

## 3. Positional Information and the Community Effect

The cells in embryos have to accomplish two feats. They need to express discrete fates in the right places in response to continuous signals. This pattern formation process is naturally broken into 'positional information' a term coined by Lewis Wolpert (see his Developmental Biology textbook), and 'community effect', introduced by John Gurdon [8].

One way for position to regulate fate is via a secreted signal, a so-called morphogen, whose level initiates some transcriptional cascade resulting in a defined fate. If the morphogen is activating, some intracell inhibitions have to operate down stream of the primary signal to exclude the low morphogen fates from regions of high morphogen. By far the best data we have on this paradigm is in *Drosophila* from the Gregor lab at Princeton.

The situation in vertebrates is more complex. Classical experiments in *Xenopus* from Smith [9] (for Activan/Nodal) and Brivanlou [10] (for BMP), used multipotent cells obtained by dissociated the *Xenopus* animal cap prior to gastrulation. Graded levels of ligands were applied, the cells re-associated and gene expression compared against similarly timed intact embryos. A 10-20x range of concentrations elicited the full range of fates in fairly discrete bands. Hence Activin/Nodal and BMP were declared morphogens. However in contrast to Drosophila, it's difficult to imagine these morphogens as static around the time of gastrulation, and none have been directly visualized at WT levels. Furthermore the classic experiments from Smith and Brivanlou assayed expression at a convenient endpoint, and already in the mid 1990s papers from JB Gurdon showed the dynamics of morphogen interpretation was more complex than assumed[11]. (A general aside: most genetic screens normalize to an endpoint that is well removed from the time at which the gene operates; this obscures the dynamic role we believe those genes should have.) Thus one may ask whether its just morphogen levels that defines fates.

An alternative view of morphogen signaling, with almost no in-vivo data, posits that cells respond to morphogens adaptively, in analogy to E.coli chemotaxis. That is the absolute level of morphogen does not matter at all provided it is static. The transcriptional network down stream of the receptor has a fixed point independent of morphogen level (some simple examples in [12]). Then by continuity, if the adaptive system does not simply ignore the stimulus, the transcriptional output is determined by a smoothed time derivative of the input where the time scale is set by the feedbacks. While negative feedbacks at multiple levels are the norm for signaling pathways, this does not imply they are adaptive. However it's easy to imagine that position relative to an unsteady source of morphogen could be inferred from the received signal. For those inclined to information theory, there is a literature on communication via a diffusive channel, but clearly the information is in the rate of change of the signal so an adaptive receiver is called for [13]. Note from the embryo's point of view both the source and receiver can be tuned by evolution to work together to define position. There is no reason to consider the information theoretic limits on reception for a presumed source of diffusible morphogen since properties of the source may also be tuned. The classical experiments on Activin/Nodal and BMP as morphogens are completely compatible with reception by an adaptive system. An adaptive transcriptional response was demonstrated for a myogenic cell line by microfluidic control of the signal in [14], and in hESC in a preprint from the Warmflash lab.

The community effect is more mysterious since multiple mechanisms contribute. Perhaps the best understood vertebrate example is the transition from the 8-cell mouse embryo where cells are nearly equivalent to the preimplantation embryo with three distinct lineages [15]Figure 3. One should perhaps digress here and define some terms from the pre-molecular era of embryology [16]. A cell is said to be:
- Competent if its able to respond to a signal
- Specified or committed if it will assume its normal fate in the absence of further signals
- Determined if its fate is unchanged even if challenged with new signals
- Differentiated if it visibly changes its morphology or identity.

Cells in each of the three lineages in the mouse blastoderm are determined, in the above nomenclature. They will only graft into the layer from which they came, which is generally how these properties were assayed in the pre-molecular era.

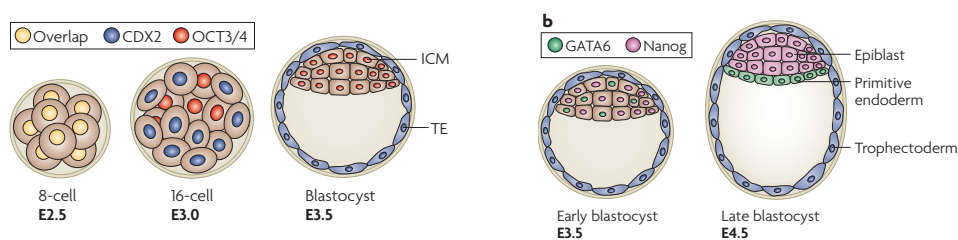

Mouse embryo:   8 cells to preimplantation

Figure 3 Schematic of mouse embryo from 8 cells to 128 cells preimplantation showing the progressive emergence of the epiblast (which gives rise to the body proper) and the extraembronic lineages, primitive endoderm and trophectoderm , along with some of the distinguishing markers [15] Fig. 1.

At the 8 cell stage the embryo 'compacts' and the cells acquire an basal (in) and apical (out) polarity [17]. By a combination of oriented cell divisions, mechanics [18], and probably mutual inhibition at the transcriptional level, the trophoblast separates from the inner cell mass ([19] and recent papers from the J. Rossant lab). A second stage of transcriptional bistability mediates the splitting of the inner cell mass. A combination of cell sorting (analogous to phase separation driven by surface tension differences between the cells) and potentially chemotaxis driven by FGF4, separates the epiblast from the primitive endoderm, [20] (and earlier papers from the Hadjantonakis lab). Finally there are isolated examples of cell death driven by cell competition, a still mysterious process at the molecular level whereby minority cells are eliminated. Thus all imaginable mechanisms contribute to lineage separation in the mouse blastula.

Hypothetically a reaction diffusion system with a nonlinear self activation of one species and its inhibition by a second activated species with a larger diffusion constant could convert a mixture of cells to two pure populations [21]. These are also the ingredients for a Turing system, and with suitable nonlinear saturation it will give rise to two discrete phases. Evidence for cooperative fate determination in a small hESC system was provided in [22], without elucidating all the molecular players.

## 4. Signaling pathways are reused

In spite of what one might read in a textbook, signaling pathways do not work in isolation in the vertebrate embryo. There is a cascade from BMP to Wnt to Nodal in the mouse that initiates primitive streak formation [15], and the same chain of induction in hESC ([23] and to appear), with similar consequences. The neural crest delaminates from the neural plate before it closes and under the control of BMP, Wnt, FGFs cells stream out and reconstitute mesoderm derivatives (bone, muscle, cartilage) and ectoderm derivatives (peripheral nerves, melanocytes). They play a major role in the morphogenesis of the vertebrate face.

The dorsal-ventral axis of the neural tube is defined by Sonic hedgehog (Shh) from the notochord and floor plate (ventral) and BMP4 from the roof plate (dorsal). Somites form in a head to tail sequence mediated by a retinoic acid gradient anteriorly and a Wnt, FGF gradient posteriorly [24]. (The A. Aulehla lab has developed somite-forming explants as an interesting model for spatial patterning.) In-vivo, the somites first condense as epithelial balls by a mesenchyml to epithelial transition (MET), which can also occur ectopically [25]. They subsequently undergo an EMT on their medial-ventral side and wrap around the spinal cord to form the vertebrae, cartilage and a second population shifts by half a period and makes the skeletal muscle that bridges the vertebrae. All these gymnastics are under the control of BMP, Wnt, Shh and their inhibitors coming from three directions: the dorsal-ventral sides of the central body axis (neural tube and notochord) and lateral mesoderm [26].

Wieschaus remarked that much of morphogenesis is like origami, the folding of epithelial sheets, but morphogenesis also entails a back and forth between the mesenchyml and the epithelial state. The transition from the presomitic mesoderm, to the somites, and back to the mobile precursors of bone and cartilege is a good example. Is this in part a mechanism to enforce discrete fates on a continuum of cells? Certainly the HOX genes must be expressed in registry with the discrete somites [27].

The point of this jumble of jargon is to delineate a broad question in the spirit of the Solvay conferences. Biologists do not ask why certain pathways are deployed in certain contexts and in certain combinations, it's too easy to rationalize it all as evolutionary artifact. The literature abounds in just-so stories, none as entertaining as Kipling. What more can be done? There is almost no biophysical and dynamical characterization of the canonical signaling pathways in an embryonic context. Are they simple ON/OFF switches, because it's assumed that disconnected cells on a dish properly report pathway response? But this ignores the fact much of development involves epithelial layers that may be apically-basally polarized. A polarized epithelium could control the reception of activators and inhibitors [23], but almost nothing is known in-vivo. To a first approximation, the embryo is still conceived as empty space where any signal can go wherever it's needed. The practical or engineering reason to address the 'why' question is that it may yield a useful phenomenological descriptions of the interrelated processes of morphogenesis and fate determination. These can be fit to data and become predictive. Even half correct theory, that really addressed global questions of pattern formation with molecular details, would greatly accelerate progress in embryology and regenerative medicine.

## 5. Stem cell biology

This subject is practically infinite, and the next three short sections serve just to delineate some concepts and open questions for a quantitative audience and provide a few key references. The subject incites a gold rush fervor with a concomitant inattention to detail, since commercial applications beckon, but in my view the best work remains well grounded in developmental biology [28],[29].

### *5.1 Organoids*

One of the most spectacular examples of organoids, and indeed the first, are the mini-guts of Sato and Clevers [30]. The human gut has a surface area of several hundred square meters, formed by a meshwork of protrusions, villi, that continuously turn over. There are specialized stem cells that occupy the base of the crypts, the slender cavities that are mixed among the villi. They self renew, and their descendants include all the more specialized cells that populate the villi. The Clevers lab found molecular markers for these stem cells and to prove their regenerative capacity they cultured isolated cells in a 3D matrix. To their surprise they made mini-guts with crypts and villi! Furthermore stem cells isolated from the mini-guts would repeat the generation process indefinitely. This system is simple enough that molecular pathways can be dissected, e.g., [31], and the usual players are at work, Wnt, EGF, Notch in the crypt, opposed by a BMP gradient from the villi. The last triumph of this system is medical [32]. Here mini-guts made from a patient biopsy were used to screen approved drugs against a rare mutation in the cystic fibrosis gene. The patient improved within hours after receiving the screening candidate.

Working from both mouse and human ESC, the J.M. Wells and J.R. Spence labs have created embryonic gut and stomach. The H. Snoeck lab has created embryonic lungs, and A. Grapin-Botton grows pancreas from stem cells. All these systems beg for quantitative modeling.

More dramatic to the public at least than these endoderm derivatives, are the optic-cups from the Sasai lab [33]from hESC, following their work in mouse. One begins from a ball of cells, it invaginations from the surface and after several additional weeks, six types of neural retinal cells form in appropriate configuration with plausible connections.

There is not yet a full convergence of groups studying mammalian embryonic development and those recapitulating parts of the process with ESC. But organoid systems for the quantitatively minded, are the best compromise between reality and tractability to study the relation of morphogenesis and differentiation. They realize the mantra 'if you can built it you understand it'.

*5.2 Adult stem cell niches*

Systems that renew routinely such as blood, the immune system, skin as well as those that renew upon injury, such as skeletal muscle, or the liver all have dedicated populations of so called adult stem cells that can recreate the necessary tissue. Typically these cells reside in compartments distinguished by structure and accompanied by specialized signaling, always involving the canonical pathways we know from development plus perhaps some specialized growth factors. Hence the question, can these niches be understood, or better predicted from what we know about development? Some prominent biologists in the field would say no.

Of particular interest in this regard are stem cell niches that can be reconstituted in-vitro, the mini-guts mentioned above being a prominent example. A second case is the satellite cells that regenerate the myotubes of skeletal muscle. They are normally are dispersed among the myotubes and not in any obvious specialized structures. The entire system of stem cells and myotubes was recreated from ESC in [34], and the functionality of the satellite cells verified by grafting.

For the systems that can only be studied in-vivo, haematopoiesis, is perhaps the most challenging since the relevant stem cells constitute of order 0.01% of the bone marrow (S. Morrison 9/28/2017 Rockefeller lecture) and reside in a structurally complex environment [35]. The medical implications of preparing haematopoietic stem cells would be immense if the homing problem could also be solved, i.e., how to get them in the right place. One could potentially cure all blood/immune cancers. Another niche studied by the E. Fuchs lab at Rockefeller is for hair cells, and the signals are Wnt, Shh, and BMP. Finally this volume has a commentary from BD Simons on branching morphogenesis in the kidney where the stem cell niche resides at the tips of the growing endothelial networks.

*5.3 Micropattern culture of hESC*

A group of physics postdocs and students working jointly with me and Ali Brivanlou at Rockefeller are exploiting hESC to recapitulate the earliest steps of embryonic patterning. Stem cells differentiated on a slide with canonical morphogens assume a variety of fates in a spatially disorganized fashion. Early endoderm protocols tolerated a lot of death but still generated useful numbers of cells for subsequent assembly steps (papers from Wells and Spence labs noted above). Our primary discovery was that mere spatial confinement in 2D micropatterned colonies induced the cells to self pattern in a reproducible way [36]. Thus cells communicated with each other in preference to the primary morphogen that was manifestly uniform in the solution. The following paragraphs summarize some results from these systems, most in the process of publication, with an emphasis on technique. The potential of these systems in reviewed in [37].

The micropatterns are 0.5-1mm in diameter and display four fates in a radially symmetric pattern that from outside to center correspond to: extraembryonic, endoderm, mesoderm, and endoderm. Their order matches that derived by projecting the cup shaped mouse epiblast onto a disk (P. Tam in [38]). The mes-endoderm cells plausibly arise by gastrulation for which both the morphogenic movements and molecular markers correspond to what we expect from the (mouse) embryo, though nothing is known molecularly about human gastrulation and only a little from non-human primates. The ~2000 cells in each micropattern define their fate by distance from the colony boundary, as shown by comparing disks of different radius. As the size shrinks, the inner fates disappear and the outer territories retain their dimensions. The same secondary morphogens and secreted inhibitors operate on the micropatterns as in the embryo.

A second paper, [23], clarified in molecular terms how cells sensed the colony edge and measured distance from it. The pluripotent colonies are apical-basal polarized epithelia, and they restrict their BMP and Activin/Nodal receptors to their baso-lateral side, thus rendering them inaccessible to apically supplied morphogens, except on the colony boundaries where the receptors become apically accessible. Growing cells on filters is a very clean way of distinguishing apical from basal responses. The second mechanism restricting signaling to the colony edge are secreted inhibitors that come into play when the BMP morphogen is applied, move laterally in the colony and leak out the edges. Pattern formation was examined in an exhaustive zoo of shapes, and all could be predicted from the data collected on disks plus the assumption of 2D diffusion with zero boundary conditions.

Cell lines with both activator and inhibitors under DOX control can readily be generated, as well as homozygous knock out lines for genes that are essential for pattern formation. Using filter grown colonies with sparsely seeded DOX inducible cells, its possible to watch the local spreading of both activators and inhibitors and how they interact with the same components applied selectively to the apical or basal sides of the colony. In a very natural context it's possible to dissect the influence of cell polarity on signaling.

There are live reporters for the BMP, Activin/Nodal, and WNT pathways. Thus signaling history can be related to cell fate. Patterning can be triggered with a secondary morphogen such as WNT, the same germ layer arrangement obtained with BMP stimulation, less the outer most extra embryonic ring, as one would infer from data in mouse.

Three-dimensional differentiation from balls of ESC, so called embryoid bodies (EB), is a common starting point for organoid development. The same technology has been used to explore the emergence of germ layers, but the results are not nearly as standardized as micropattern culture, imaging is more complex, and there has been far less molecular dissection of the signaling [39]. But by far the biggest problem with these systems as a model for gastrulation related events, in mammals is that an epithelial cell population, the epiblast, initiates the process and gives rise to the entire adult body.

(Incorrect inferences from EB as to how the mouse inner cell mass cavitates to form the epiblast were only corrected in [40].). Human ESC are technically an easier starting point for gastrulation since they naturally propagate in a state very analogous to the post implantation epiblast, while mouse ESC resemble the preimplantation inner cell mass. The mESC can be converted to epiblast cells, but the resulting state is not entirely stable and seems more variable than the normal hESC (details are technical).

Our own technique for work in 3D, seeds single cells in a specially tailored matrix and allows them grow into an epithelial shell with basal out and apical in while remaining pluripotent (Simunovic et.al.)`. A very gentle BMP stimulus results in spontaneous polarization of the epithelial cysts into a primitive streak region, showing all the markers of gastrulation (that define the future posterior), and a complementary region with anterior epiblast (future ectoderm) markers. This is a true symmetry breaking and does not require an asymmetric source of BMP as in prior work with mouse [41]. Another variant on this method of 3D culture even results in morphological symmetry breaking prior to gastrulation [42].

An important step in taming mESC to explore gastrulation and beyond in the mouse has been taken in a forthcoming paper by Morgani and Hadjantonakis. They devised a protocol to recreate the pre-gastrulation mouse epiblast on micropatterns. They then added the BMP and Wnt morphogens that would normally come from the extraembryonic tissues and observed radial patterning. The same antibody combinations could be applied to the mouse embryo at successive time points, and the correspondence of patterns and fates mapped. The details are too voluminous to recount here but are very encouraging. In the absence of any data from human embryos undergoing gastrulation it's essential to benchmark the micropattern technique against some embryonic system.

The technology used in the mouse micropattern paper is conventional antibody stains for triples of markers. This scores the proteins and allows co-stains for signaling effectors. Space-time specific expression data resolved down to a few cells for the mid-late streak mouse embryo can be found in [43]. Single cell RNA-seq is appealing technology but in development it needs to retain its time-space label. Not all genes deserve equal weight, the Hadjantonakis study focused on those with an interesting phenotype.

*6 Phenomenology*

The physical reader should realize this term is a pejorative in biological contexts. It denotes a return to $19^{th}$ century biology and the absence of the methods that made $20^{th}$ century biology great: genetics, biochemistry, structural biology, molecular biology etc. However the modeler who embraces these advances literally is doomed, at least in the area of development. A glance at the molecular constituents for any of the signaling pathways (e.g., Wnt homepage maintained by the Nusse lab) reveals 5-10 core constituents decorated by another 10-50 modifiers. The molecular complexity frustrates transferring actual numbers between systems, and the most common description of reactions with the Michaelis-Menten system requires many parameters. The solution sometimes adapted is to randomly sample parameters and select those behaviors obtained most frequently. This shows that random equations can do many things, but more fundamentally is contrary to the incrementalism that we believe is inherent in Darwinian evolution, unless you think that Diana sprung fully formed from the head of Zeus as depicted on ancient Greek vases.

Examples of successful phenomenology in the context of development are rare and I myopically mention some examples of mine and from my immediate collaborators. The foundation of the approach goes back to a book written by a student of Waddington, [16] and are based on translating the embryological concepts of competence, commitment, and determination to the language of dynamical systems. The necessary mathematics is embodied by the subject of Morse-Smale dynamical systems (http://www.scholarpedia.org/article/Morse-Smale_systems). Colloquially these are systems of equations whose limiting behavior both forward and backward in time are a finite set of fixed points and periodic orbits. They are rich enough to cover anything we can hope to measure and describe in developmental biology, even if we put aside the periodic orbits. Gradient flows with some technical assumptions are Morse-Smale.

The mapping between classical embryology and mathematics equates an equivalence group of cells [16] to the direct product of the model used for one cell. Commitment is, with various nuances, flow into a fixed point. The power of this brand of phenomenology, and also its point of failure is whether the parameters can be fit to encompass all available data. Typically gene knockouts, and overexpression data is available, but more informative is always dynamic interventions made while the system is poised among multiple outcomes. If a multivariable system has two stable states, then the simplest phenomenological model would replace it with the relaxational dynamics induced by a quartic potential in one dimension. Various morphogens would tilt the potential and favor one state over the other, ultimately by annihilating one fixed point with the saddle in a reverse saddle node bifurcation. It's clear how to add noise to the system (partial penetrance in genetic language), whose biological source could be environmental, epigenetics (molecular tags on DNA and chromatin that vary between animals), or true molecular noise. The problem becomes interesting if there are multiple experimental handles on the relative stability of the two states. The challenge is to represent them all in terms of the coefficients in the potential. The first guess would be a linear combination exactly parallel to what is done for computational neural nets, where a linear weighted sum of inputs is put through a nonlinear function.

While such a representation seems very antithetical to a Michaelis-Menten network it is not so far removed from development. The interesting mutations in development do not create fundamentally new structures, but rather permute known ones. Genetics is based on quantifying recognizable characters. A fried embryo is not informative, but the old observation of genetic assimilation, that environmental insults often phenocopy genetic ones is profound. Thus we suppose that evolution has added multiple layers of regulation, many still unknown, to insure the stability of the two states in the above example. Phenomenology accepts those states and concentrates on the simpler problem of parameterizing the dynamics mediated by the morphogens (during the competence period but prior to commitment) that control the competition among them.

Phenomenology becomes more interesting when three states are in play. It's informative to enumerate a hierarchy of models by enumerating critical points and their connections in various spatial dimensions, and parameterizing the vector fields to within topological equivalence. A nontrivial example for vulva development in *C.elegans* is given in [44], and a forth coming paper by the same authors in eLife. An application to intermediate range signaling by Notch-Delta was presented in [45].

Phenomenology should also be the preferred description for moving boundary problems e.g., [46]. When two locally stable states are separated by an interface and some component of the system can move between cells giving rise to defacto diffusion, then its very appealing to model it as the relaxation of a bistable free energy with a spatial gradient following Kolmogorov. A more prosaic use of phenomenology parameterized the cellular response to a morphogen and coupled it with a reaction-diffusion system for the secreted inhibitor to the nominally uniform morphogen [23].

Another avenue for phenomenological reasoning codifies the continuity of outcomes with variable morphogen levels by a phase diagram. Clearly the entire signaling history impacts the pattern of embryonic fates, and for present purposes we plot terminal outcomes as a function of signal levels imposed by genetic means. (We ignore the specifics of those genetic manipulations here, so as to succinctly illustrate the ideas.)

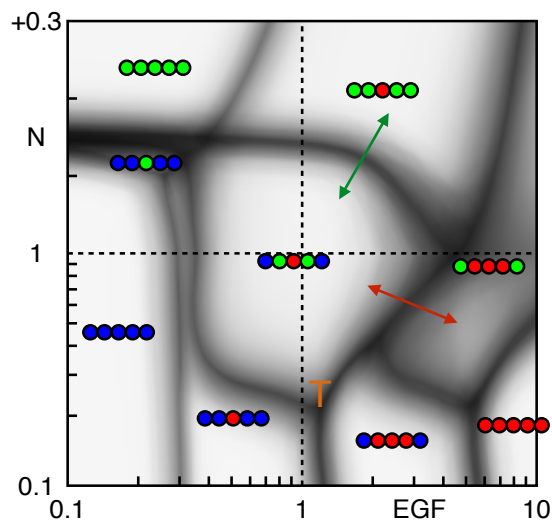

Figure 4 Phase diagram for the states of five cells each of which can assume three states (RGB) under the control of two morphogens N(otch) and EGF. Pure states are bounded by grey boundaries showing zones of mixed fates. The green arrow shows a generic transition where only one cell (reflection symmetry is imposed) changes state, the red arrow shows a correlated change and the T shows one example of a triple point where three boundaries meet (Corson & Siggia to appear).

The representation in Figure 4, which is actually computed from a model of vulva patterning, exemplifies that it would be surprising given a complex model to observe boundaries strictly parallel to the coordinate axes. Thus it's logically impossible to assert that N controls just the green fate and EGF just the red, as one might infer from a casual reading of biological papers, but rather there is simply more green on top and more red on the lower right. Once phase boundaries are freed from alignment with the coordinate axes, they generically met in triple points. These are points where conventional genetic analysis becomes complex and thus interesting. It's evident that the most informative data for fitting the dynamical model that underlies the phase diagram, are precisely those genetic backgrounds that yield mixed fates (partial penetrance in the jargon). Thus merely codifying the obvious yields insights.

The developmental geneticist uses a *sensitized* background to define the activity of a new mutant that has no effect in the wild type background. In mathematical terms the sensitized state is one near to inset to a saddle point, that is the ridge ending in the pass that separates two basins of attraction [44]. It is typically difficult to infer by verbal reasoning alone the activity of the silent mutation from the identity of the terminal states. But such data can be very useful in model fitting. More generally viewed from the perspective of modeling, the most informative experiments apply time dependent perturbations to the system while the decisions among states are being made. This modality is the complete converse to the typical genetic screen where time is eliminated and only a late terminal state is recorded. Although genetics furnishes us with a parts list for development, the description of those parts is rather removed from the context in which they function. Imposing a Morse-Smale description on development may ultimately prove to be incomplete, but in the interim, certainly suggests many informative dynamic experiments.

Can theory do more than principled data fitting in cell and developmental biology? An old article by Jacob [47] reminds us that evolution works by tinkering, rearranging existing parts. Darwin, in his oft-quoted passage on the evolution of the vertebrate eye, observes that complex structures can be created rapidly by gradient search. Models for various slices of development were derived by gradient optimization in a series of papers by P. Francois and me. Some rather nonobvious dynamical models emerged with their specific parameters and no appeal to parameter space volume. A template for how to generalize these ad-hoc simulations to general theory is provided by models in machine learning [48]. Within a defined learning environment, it is shown some rules can be learned from only a polynomial number of examples, while others require an exponential number. One would expect the evolutionary tinkerer to discover only the former class.

The biologist's aversion to phenomenology has a specific connotation in development. If genes are the atoms of biology, can a phenomenological model ever constitute fundamental understanding? However if the genetic description is infinitely complex, do we really learn anything from an equally complex model? There is a parallel debate about the uses and abuses of phenomenology in neuroscience ([49] and the Oct. 27 2017 issue of *Science*).

## 7 Perspectives

To categorize an embryo as an instance of non-equilibrium symmetry breaking, reduces embryology to banal physical categories that hide the interesting phenomena. In the words of C. H. Waddington in his 1966 Principles of Development and Differentiation

> *"To anyone with his normal quota of curiosity, developing embryos are perhaps the most intriguing objects that nature has to offer. If you look at one quite simply .... and without preconceptions .... what you see is a simple lump of jelly that .... begins changing in shape and texture, developing new parts, sticking out processes, folding up in some regions and spreading out in others, until it eventually turns into a recognizable small plant or worm or insect...*
>
> *Nothing else that one can see puts on a performance which is both so apparently simple and spontaneous and yet, when you think about it, so mysterious."*

While no one believes that new chemistry or physics is required to treat biology, some thought is necessary to arrive at an informative level of description, just as in neuroscience, [49].

Biological literature is often difficult to penetrate by the non-expert, since in contrast to physics there is less a tendency to publically rebut dubious results. People in the field know, but those on the outside do not, absent private discussions. A welcome exception to that norm, and one I was chided for not mentioning by Prof Alon, is a paper from the Barkai group on scaling in Xenopus [7]. A paper from prominent *Xenopus*, and stem cell biologist, Y. Sasai [50], notes on p1308 "experimental observation in the present and previous studies do not appear to support … this model" [51]. Several quibbles were also raised on theoretical grounds to [7]: the gastrulating embryo is not one dimensional, the theory has of order 30 parameters to fit almost no data, and a generic reaction-diffusion model will account for the scaling of half sized embryos as well [52] (also noted in [51]). Happily in this instance, science was self-correcting.

Turing patterns are invoked in a wide variety of contexts. I prefer to use the term in the strict sense of a reaction diffusion system that leads to spontaneous spatial pattern with a wavelength determined by diffusion constants and reaction rates. The regulatory system that controls the three axes of the vertebrate limb, proximal distal, AP (thumb is anterior) and DV (palm is ventral) has a long history in embryology [53], and the amphibian limb is currently a key model system for the molecular understanding of regeneration (E. Tanaka lab). Turing physics has been an appealing explanation of vertebrate digits since their periodicity can be decoupled from their identity [54]. However a very instructive rebuttal to molecular data for a Turing origin of the vertebrate digits [55] was given in [54]. Alternative models of periodic patterns may involve mechanics [56].

The reason for concluding this opinion piece with examples with flawed models of developmental systems published in visible forums, is to impress upon the reader the diversity of facts that can impinge upon a model and desirability to partner with a lab conversant with those facts. Failed models are a sign of progress, since data and theory are engaged.

**Acknowledgement:** This work was supported by NSF grant PHY 1502151.